# Strong Pseudospin-Lattice Coupling in $Sr_3Ir_2O_7$: Coherent Phonon Anomaly and Negative Thermal Expansion

L. L. Hu[1], M. Yang[1,6], Y. L. Wu[1], Q. Wu[1,6], H. Zhao[1,6], F. Sun[1,6], W. Wang[2], Rui He[3], S. L. He[4], H. Zhang[5], R. J. Huang[2], L. F. Li[2], Y. G. Shi[1,6], Jimin Zhao[1,6,7,]*

[1] *Beijing National Laboratory for Condensed Matter Physics, Institute of Physics, Chinese Academy of Sciences, Beijing 100190, China*

[2] *Technical Institute of Physics and Chemistry, Chinese Academy of Sciences, Beijing 100190, China*

[3] *Department of Electrical and Computer Engineering, Texas Tech University, Lubbock, TX 79409, USA*

[4] *Ningbo Institute of Material Technology & Engineering, Chinese Academy of Sciences, Ningbo 315201, China*

[5] *College of Physical Science and Technology, Sichuan University, Chengdu 610065, China*

[6] *School of Physical Sciences, University of Chinese Academy of Sciences, Beijing 100049, China*

[7] *School of Physical Sciences, University of Chinese Academy of Sciences, Beijing 100049, China*

* Correspondence and request of materials should be addressed to Jimin Zhao. Email: jmzhao@iphy.ac.cn

# Abstract

The similarities to cuprates make iridates an interesting potential platform for investigating superconductivity. Equally attractive are their puzzling complex intrinsic interactions. Here, we report an ultrafast optical spectroscopy investigation of a coherent phonon mode in $Sr_3Ir_2O_7$, a bilayer Ruddlesden-Popper perovskite iridate. An anomaly in the $A_{1g}$ optical phonon (ν = 4.4 THz) is unambiguously observed below the Néel temperature ($T_N$), which we attribute to pseudospin-lattice coupling (PLC). Significantly, we find that PLC is the dominant interaction at low temperature, and we directly measure the PLC coefficient to be $\lambda = 150 \pm 20$ cm$^{-1}$, which is two orders of magnitude higher than that in manganites (< 2.4 cm$^{-1}$) and comparable to that in CuO (50 cm$^{-1}$, the strongest PLC or spin-lattice coupling (SLC) previously known). Moreover, we find that the strong PLC induces an anisotropic negative thermal expansion. Our findings highlight the key role of PLC in iridates and uncovers another intriguing similarity to cuprates.

# I. INTRODUCTION

Single-layer and bilayer Ruddlesden-Popper perovskite iridates ($Sr_{n+1}Ir_nO_{3n+1}$, n = 1, 2 and ∞) are highly analogous to the superconducting cuprates [1,2]. The prospect that superconductivity could be induced in iridates [1] makes investigation of the complex interactions among their various degrees of freedom [3,4] particularly interesting. However, so far the major interactions in the iridates have not been fully elucidated, and for the known interactions, the physical picture is hitherto not entirely clear. Here we carry out ultrafast optical spectroscopy studies of the lattice degree of freedom in a $Sr_3Ir_2O_7$ single crystal. We directly measure the PLC strength and demonstrate that it is an essential interaction at low temperature. The strong PLC causes a phonon anomaly and leads to enhanced interlayer spacing—a negative thermal expansion.

As a typical 5$d$ compound, $Sr_3Ir_2O_7$ experiences strong spin-orbit coupling (SOC), which renders pseudospin $J$, instead of spin $S$ or orbital angular momentum $L$, a good quantum number. The pseudospins (with $J = 1/2$) align along the $c$-axis into antiferromagnetic (AFM) long-range order, whereby the AFM to paramagnetic (PM) phase transition occurs at $T_N = 280$ K [5]. It was reported that $Sr_3Ir_2O_7$ also exhibits a metal-insulator transition at $T_N$, thus suggesting $T_N = T_{MIT}$ [6]. Pseudospin has been previously proposed and investigated in iridates in Ref. [7-9], along with some pioneering discussions and fruitful conclusions on its interaction with lattice. So far, below $T_N$, PLC has not been thoroughly investigated and no clear physical picture has been obtained about the complex interactions. For example, below $T_N$, it is unknown what the dominant interaction among the multiple degrees of freedom is, whether and how the lattice interacts with the pseudospin, *etc*.

Ultrafast optical spectroscopy is a powerful method to investigate the coherent phonons [10-12], squeezed magnons [13,14], and excited-state carrier dynamics [10,15,16], as well as laser-induced electron coherence [17] and superconducting phenomena [18] in quantum materials [19]. Despite of this, ultrafast spectroscopy has not been utilized to investigate the PLC. There are only a few time-resolved ultrafast dynamics investigations of perovskite iridates reported so far [6,15], and furthermore, thorough investigations of coherent phonons are still lacking. For instance, coherent phonons in the PM metallic phase (above $T_N = T_{MIT}$) of iridates, including $Sr_3Ir_2O_7$, have not been successfully detected; and phonon anomaly in the AFM insulating phase (below $T_N = T_{MIT}$) has not been thoroughly investigated.

Here we report a thorough ultrafast optical spectroscopy investigation of the coherent phonon in $Sr_3Ir_2O_7$. We coherently generate and detect the low lying $A_{1g}$ phonon. We unambiguously observe coherent phonons above $T_N = T_{MIT}$ and, more importantly, discover a phonon anomaly below $T_N = T_{MIT}$. We attribute this phonon anomaly to a coupling between the *c*-axis AFM pseudospin long-range order and the out-of-plane breathing $A_{1g}$ phonon mode (*i.e.*, PLC). Significantly, we obtain the PLC coefficient $\lambda = 150 \pm 20$ cm$^{-1}$, which is greater than the highest values ever reported for PLC or SLC, and we identify PLC as a major intrinsic interaction in $Sr_3Ir_2O_7$ at low temperatures. Moreover, we infer from the phonon anomaly that the interlayer spacing is expanded below the Néel temperature, which is verified by our temperature-dependent x-ray diffraction (XRD) experiment. Our unambiguous experimental evidences thus paint a clear picture of the PLC in $Sr_3Ir_2O_7$.

## II. EXPERIMENT

Our pump-probe experiment is implemented in a way similar to that in Ref. [10]. Ultrafast laser pulses of 800 nm central wavelength, 70 fs pulse duration, and 250 kHz repetition rate are used to generate and detect coherent phonons in $Sr_3Ir_2O_7$. The 1.55 eV photons excite the transitions from the $J_{eff} = 3/2$ valence band to the $J_{eff} = 1/2$ conduction band [20] (see Appendix A). The polarizations of the pump and probe laser beams are chosen to be perpendicular to each other to minimize the noise. We measure the differential reflectivity ΔR/R as a function of the delay time of the probe beam. The pump and probe fluences are maintained at 0.68 mJ/cm$^2$ and 0.46 mJ/cm$^2$, respectively. The diameter of the focal spot for the pump and probe beams are 42 μm and 33 μm, respectively. The pump beam is modulated at 2 kHz by a mechanical chopper. The reflected probe beam is collected using a balanced photo-detector and sampled by a lock-in amplifier to enhance the signal-to-noise ratio. Temperature dependence experiments are performed with an open cycle liquid Helium cryostat.

Our temperature-dependent XRD experiment on $Sr_3Ir_2O_7$ single crystal is carried out using a BRUKER Discover D8 diffractometer from 20 to 310 K, with the Cu Kα radiations (with wavelengths of Kα$_1$ and Kα$_2$ being 1.5406 Å and 1.5444 Å, respectively). At each temperature, the 2θ angle is varied and recorded from 10 to 60 degrees, with a step increment of 0.001 degree. The XRD patterns are indexed in the space group $I4/mmm$. From the recorded 2θ diffraction peaks, the lattice parameter is calculated using Bragg's law.

We first grow a single crystal of $Sr_3Ir_2O_7$ (see Appendix B) and then we use Raman scattering and single crystal XRD (see Appendix C) to characterize the sample. The Raman scattering is consistent with the reported Raman spectra of $Sr_3Ir_2O_7$ [7]. By comparison with the standard XRD results, we assign the cleaved surface of our $Sr_3Ir_2O_7$ sample to be the (0 0 1) plane (see Appendix C).

## III. RESULTS AND ANALYSIS

The ultrafast dynamics result at 6 K is shown in Fig. 1, with the background scattering noise subtracted. The scanning trace is composed of exponential decays (the electronic dynamics) and superimposed periodic oscillations (the coherent phonon dynamics). After subtracting the electronic relaxation, the damped oscillation is re-plotted in the upper inset of Fig. 1. We attribute the oscillation to a coherent optical phonon mode. The fast Fourier transformation (FFT) analysis yields a phonon frequency of 4.42 THz (*i.e.*, 147 cm$^{-1}$), as explicitly shown in the lower inset of Fig. 1. By comparing with the Raman results (see Appendix C), we assign the coherent phonon to be the lower-frequency $A_{1g}$ mode. In the FFT result, there is also a discernible peak at 5.5 THz (*i.e.*, 183 cm$^{-1}$), which corresponds to the higher frequency $A_{1g}$ phonon. In the upper inset, the fitting curves of both phonon modes are plotted, respectively. In the following we only discuss the temperature dependence of the lower frequency $A_{1g}$ phonon (red curve). It can be seen that the dynamics of this coherent phonon mode exhibits a cos($\omega t$) time dependence, with an initial phase of zero (see the upper inset of Fig. 1 and Fig. 2(a)). Thus, we attribute the generation mechanism of this oscillation to be displacive excitation of coherent phonons (DECP) [21-23]. The lower frequency $A_{1g}$ phonon corresponds to the out-of-plane breathing vibration of the iridium oxide bilayer. The relevant $IrO_6$ octahedra are centered on adjacent $Ir^{4+}$ ions, carrying opposite pseudospins oriented along the $c$-axis (see Appendix A). Note that the low-lying $A_{1g}$ coherent phonon has a much larger magnitude than that of the higher energy $A_{1g}$ coherent phonon (inset of Fig. 1), which is unlike the case observed in Raman scattering [7]. This is mainly because the excitation laser wavelength in Ref. [7] was 633 nm, and the one we use here is 800 nm (see Appendix E).

We further investigate the temperature dependence of the coherent phonon in the range of 6-320 K (Fig. 2). In Fig. 2(a) a few typical scanning traces are shown, which are offset for clarity. More complete data at different temperatures are illustrated in a three-dimensional colormap (Fig. 2(b)) and a two-dimensional colormap (Fig. 2(c)). It can be seen from Fig. 2(b) that both amplitude and lifetime of the coherent phonon vary systematically with temperature. In Fig. 2(c), the phonon oscillation is illustrated explicitly, exhibiting clear phases of the coherent phonon. We fit it with

$$\Delta R/R = A_\Omega \exp(-t/\tau_\Omega)\cos(\Omega t+\varphi), \qquad (1)$$

where $A_\Omega$ is the phonon amplitude, $\Omega$ is the phonon frequency, $\tau_\Omega$ is the phonon lifetime, and $\varphi$ is the initial phase. These coherent phonon parameters are fitted for different temperatures, and are plotted in Fig. 3. We discuss the underlying physics as follows.

First, the phonon amplitude $A_\Omega$ exhibits an anomaly at and below $T_N$, as shown in Fig. 3(a). For decreasing temperature, the phonon amplitude (black curve) keeps nearly constant from 320 K to $T_N$ = 280 K (in all the figures, $T_N$ is marked by a vertical dashed line). Then it suddenly increases dramatically from $T_N$ = 280 K to 180 K. After that, the increase in amplitude gradually slows down, until reaching the liquid Helium temperature. As seen, the slope of $A_\Omega$ clearly exhibits a sharp break across $T_N$. The abrupt enhancement upon the onset of the AFM ordering unambiguously demonstrates the coupling between this phonon mode and the pseudospin AFM order. A 5-fold enhancement in $A_\Omega$ on entering the ordered pseudospin state (from the metallic PM state to the insulating AFM state) is identified (see Figs. 2 and 3(a)). The spiking increase of amplitude triggered by the AFM order below $T_N$ cannot be solely explained by a regular thermal expansion (see Fig. 1 of Ref. [11]); thus, we identify this phenomenon as a *phonon anomaly*. The blue dashed curve in Fig. 3(a) represents the AFM order parameter ($M_{AF}^2$) adapted from Ref. [24], which vanishes above $T_N$. It can be seen that $A_\Omega$ exhibits a similar temperature dependence to that of $M_{AF}^2$. To our knowledge, such phonon behavior has not been reported before for 5$d$ compounds. The strong similarity in temperature dependence between the two quantities suggests that the coherent phonon is correlated to the collinear long-range AFM order below $T_N$; as such, PLC must exist below $T_N$, which directly leads to the observed phonon anomaly.

We then quantitatively understand this PLC-induced phonon anomaly in amplitude. It is known that, for coherent phonons, the amplitude $A_\Omega$ is linearly dependent on the Raman tensor $\partial\alpha/\partial Q$ [21], where $\alpha$ is the polarizability and $Q$ is the normal coordinate of the phonon mode. The integrated intensity $I(T)$ of phonon Raman scattering in a magnetic material has been phenomenologically described [25] as $I(T) = \beta |Z + \Lambda <S_i \cdot S_{i+1}>|^2$, where $Z$ represents the spin-independent term, $\Lambda$ represents the spin-dependent term, $<S_i \cdot S_{i+1}>$ is the correlation function between two nearest neighboring spins, and $\beta$ is a constant. Microscopic analysis showed that the term containing $\Lambda$ is mainly contributed by the $d$ electron hopping between nearest neighboring magnetic ions during phonon Raman scattering [25]. In our case, it reflects the coupling between the pseudospin and the phonon. The term $<S_i \cdot S_{i+1}>$ can be replaced by $<J_i \cdot J_{i+1}>$ owing to the strong SOC. Because the Raman tensor is proportional to $\sqrt{I(T)}$, we have $A_\Omega \propto \partial\alpha/\partial Q \propto \sqrt{I(T)} \propto Z + \Lambda <J_i \cdot J_{i+1}>$ in Sr$_3$Ir$_2$O$_7$, where $J_i$ and $J_{i+1}$ are pseudospins of the nearest neighboring Ir$^{4+}$ ions. In the simple Heisenberg nearest neighbor approximation and the molecular field approximation, the pseudospin correlation is given by $<J_i \cdot J_{i+1}> \approx -M_{AF}^2/4\mu_B^2$ (see Appendix D). Thus we have

$$A_\Omega \propto Z - \left(\Lambda/4\mu_B^2\right) M_{AF}^2. \tag{2}$$

This correlation is a direct manifestation of the PLC. It can be seen that this relation is directly exhibited in the experimental results shown in Fig. 3(a). The temperature dependence of $A_\Omega$ is in line with the $Z/\Lambda < 0$ case in Ref. [25]. Moreover, in Fig. 3(a), there is a non-zero residual above $T_N$, which we attribute to the contribution of the pseudospin-independent term $Z$. Overall our experimental results compare very well with the theoretical expectations. The proportionality between $A_\Omega$ and $M_{AF}^2$ has not been reported before, which strongly indicates that PLC is a crucial interaction in Sr$_3$Ir$_2$O$_7$ and the phonon amplitude anomaly is a direct effect of the PLC.

Second, the phonon frequency also exhibits an anomaly below $T_N$ (Figs. 3(b) and (c)). In Fig. 3(b), we plot the temperature dependence of the phonon frequency (red curve). The slope of the frequency curve experiences a sharp kink at $T_N$, which cannot be explained by regular anharmonic phonon interactions (*i.e.*, thermal expansion; compare with the inset of Fig. 3 in Ref. [11]). To see this explicitly, we plot regular anharmonic phonon-phonon scattering behavior in the same figure: we fit the data points at temperatures $T \geq T_N$ by using the anharmonic interaction model $\Omega(T)=\Omega_0+C(1+\frac{2}{\exp(\hbar\Omega_0/2k_BT)-1})$, where $\Omega_0$ and $C$ are constants [26]. The fitting result is depicted in Fig. 3(b) by a black curve (dashed for $T < 280$ K and solid for $T \geq 280$ K). It can be seen that, for $T < 280$ K, a prominent deviation in frequency is observed (shaded area) between the experimental results and the anharmonic scattering expectation. This deviation is unusual for a regular solid, which indicates that extra forms of interaction exist and lower the phonon frequency for $T < T_N$. We subtract the experimental frequency data from the expected curve of anharmonic scattering to obtain $\Delta v$ (*i.e.*, $\Delta\Omega/2\pi$), which is re-plotted in Fig. 3(c). The temperature dependence of $M_{AF}^2$ is again adapted from Ref. [24]. The magnitudes of $\Delta v$ and $M_{AF}^2$ show similar temperature dependence, especially for low temperatures. The similar temperature dependence between the two quantities strongly indicates that PLC is a significant interaction in Sr$_3$Ir$_2$O$_7$. Such a finding again clearly indicates that the phonon frequency anomaly we observe is also a direct result of the PLC.

Quantitatively, the PLC leads to a frequency shift of [27]:

$$\Delta\Omega = -\frac{1}{8\mu\Omega}\frac{\partial^2 J_c}{\partial u^2}\left(\frac{M_{AF}}{\mu_B}\right)^2, \tag{3}$$

where $\mu_B$ is the Bohr magneton, $\mu$ is the reduced mass, $\Omega$ is the phonon frequency at 0 K (in Fig. 3(b) we obtain $\Omega/2\pi$ = 4.420 ± 0.007 THz (*i.e.*, 147.3 ± 0.2 cm$^{-1}$)), and $\frac{\partial^2 J_c}{\partial u^2}$ is the phonon modulation of the super-exchange interaction $J_c$ along the *c*-axis.

The quantity $\frac{\partial^2 J_c}{\partial u^2}$ is proportional to the PLC coefficient $\lambda$, with

$$\lambda = -4\Delta\nu\mu_B^2/M_{AF}^2 = \frac{1}{4\pi\mu\Omega}\frac{\partial^2 J_c}{\partial u^2}$$ 

([28], see Eq. (9) and (12) in Appendix D). Based on our experimental results shown in Fig. 3(c), by considering the low temperature regime with prominent AFM ordering, we directly measure $\Delta\nu$ = -0.15 ± 0.02 THz (*i.e.*, -5.0 ± 0.6 cm$^{-1}$) (with $M_{AF}^2$ = 0.133 $\mu_B^2$ [24]), by which we obtain the PLC coefficient $\lambda$ = 150 ± 20 cm$^{-1}$ (*i.e.*, $\frac{\partial^2 J_c}{\partial u^2}$ = 1160 ± 154 *m*Ry/Å$^2$, see Appendix D). This $\lambda$ value is significantly greater than those reported values of other materials (Table 1). For example, the value of SLC coefficient $\lambda$ for LaMnO$_3$ is 2.4 cm$^{-1}$ [27], and the value of SLC coefficient $\lambda$ for CuO is 50 cm$^{-1}$ [28], which was the largest $\lambda$ value previously reported. We attribute the exceptionally large value of the PLC coefficient $\lambda$ to the relatively stronger SOC in 5*d* compounds. Because the spins are intertwined with the orbitals by SOC, and orbitals are naturally coupled to the lattice, the resulting pseudospins also couple strongly to the lattice.

Third, we analyze the temperature dependence of the phonon lifetime, which also exhibits some anomalous features. In Fig. 3(d) the phonon lifetime (blue curve) is nearly constant at above $T_N$, and it starts to increase when encountering the onset of the long range pseudospin ordering at $T_N$ and upon further cooling. We plot the reported temperature dependence of the resistivity on a logarithmic scale [32] in the same figure (red dashed curve). Interestingly, its temperature dependence exhibits similar features as the lifetime. The regular anharmonic phonon interaction picture again cannot fully explain the observed lifetime. Microscopically, upon cooling, a Mott insulator releases less itinerant thermal electrons. The effect of cooling is two-fold: it increases the phonon lifetime because less thermal carriers involve in the phonon-carrier scattering, and it increases the electronic resistivity because less thermal carriers involve in conducting currents. This explains the similarities between the curves in Fig. 3(d)—both reflect stronger electron-phonon interactions at higher temperatures. We attribute this lifetime anomaly to a direct consequence of the electron-phonon (*i.e.*, charge-lattice) interaction, which has also been observed in other investigations [4].

Finally, we consider the exact physical picture of the PLC observed in Fig. 3(a)-(c). We have ascribed both the amplitude and frequency anomalies as effects of PLC. As a general consideration, upon cooling, the frequency reduction $\Delta v$ (Fig. 3(b) and (c)) must be caused by reduced total layer-layer interaction. Since $\Delta v$ is proportional to $M_{AF}^2$ at low temperatures, we attribute this reduction of interlayer interactions to the PLC. Classically, the PLC acts like a "spring" with negative spring constant in addition to the regular lattice interaction. For regular lattice interactions (regular "springs"), cooling leads to a reduced interlayer spacing. Thus the contribution from a "negative spring" will lead to expanded interlayer spacing instead.

In Fig. 3(e) we schematically illustrate this picture, where in the low temperature AFM state the interlayer spacing is expanded. Importantly, this property is opposite to the regular thermal expansion picture. Thus our experimental results predict a novel consequence of PLC in $Sr_3Ir_2O_7$—a negative thermal expansion along the *c*-axis. Furthermore, as seen from Fig. 3(e), an enhanced interlayer spacing reduces the overlap (hybridization) between the *d* orbitals of the neighboring $Ir^{4+}$ ions and the *p* orbitals of the bridging $O^{2-}$ ions, leading to reduced orbital interactions in $Sr_3Ir_2O_7$.

To further verify the above interpretations obtained from our ultrafast spectroscopy investigation, we carry out a temperature-dependent XRD experiment to directly measure the lattice constant *c* (hence the $IrO_6$ interlayer spacing) of the $Sr_3Ir_2O_7$ single crystal. We show the results in Fig. 4. The colormap of the XRD scattering intensity is presented in Fig. 4(a), where the (0 0 10) peaks are recorded at different temperatures. Their continuous and smooth evolution indicates that no structural phase transition occurs in the 20-310 K temperature range, which is consistent with Ref. [5]. From the XRD results in Fig. 4(a), we obtain the lattice constant *c* from the $2\theta$ angle value of the (0 0 10) peak. Figure 4(b) shows the temperature dependence of *c* (dark green dots). Also depicted is the expected behavior of *c* assuming linear thermal expansion (black dashed line, fitted using the $T > T_N$ experimental data). Notably, a clear discrepancy occurs in the AFM regime, where the experimental results deviate prominently from linear thermal expansion. A closer examination shows that this deviation is an increase of lattice constant *c* that begins when entering the AFM regime. We then subtract the expected linear expansion values from the experimental results in Fig. 4(b), and plot the difference $\Delta c$ (brown squares) in Fig. 4(c). The value of $M_{AF}^2$ [24] is also included (blue dashed curve). Clearly, $\Delta c$ and $M_{AF}^2$ show similar trend of temperature dependence and, except for a deviation area we discuss in a latter paragraph, they show certain proportionality to each other. Since the change in the lattice constant $c$ reflects also the change in the IrO$_6$ interlayer distance $d$, our XRD results directly reflect the PLC-induced negative thermal expansion depicted in Fig. 3(e). Because the lattice constant $a$ shows typical thermal expansion [33], the negative thermal expansion induced by the strong PLC is indeed anisotropic. Our temperature-dependent XRD result confirms the prediction of the PLC-induced negative thermal expansion based on our ultrafast optical spectroscopy results, which strongly supports the crucial importance of PLC in Sr$_3$Ir$_2$O$_7$. We note that the PLC we observe here is different from the magnetoelastic locking observed earlier in the single-layer counterpart Sr$_2$IrO$_4$ [8,9]. For Sr$_3$Ir$_2$O$_7$, the lattice expansion along $c$-axis that first revealed here only exists below $T_N$; while for Sr$_2$IrO$_4$, the lattice rotation along the $c$-axis and the IrO$_6$ octahedra distortion persists both below and above $T_N$ [9]. The PLC we observe here is in parallel to the magnetoelastic locking [8,9], and the latter is indeed also a different type of PLC.

## IV. DISCUSSION

In analyzing the amplitude anomaly, it is worthy to note that resonant Raman scattering has also been proposed to explain the temperature-dependent Raman intensity. In a different material HgCr$_2$Se$_4$, the electronic structure varies with temperature and the spin-dependent Raman mechanism does not apply satisfactorily [34]. Here for $Sr_3Ir_2O_7$, the $A_\Omega$ ($\propto \partial\alpha/\partial Q \propto \sqrt{I(T)}$) has good proportionality to $M_{AF}^2$, which was not found for $HgCr_2Se_4$ [34]. Hence we attribute the temperature dependence of $A_\Omega$ for $Sr_3Ir_2O_7$ mainly to the PLC. Furthermore, Coulomb screening may also form an alternative explanation for the observed phonon anomaly. However this mechanism strongly relies on thermally excited itinerant carriers. This will lead to the most pronounced effect nearby $T_N$, or even proceeding to the metallic state above $T_N$, which contradicts the experimental facts we observe here.

Furthermore, the overall similar temperature dependence between the phonon amplitude (Fig. 3(a)), the frequency change (Figs. 3(b)&(c))) and the lattice parameter change (Figs. 4(b)&(c)) and the magnetic order parameter $M_{AF}^2$ demonstrates that strong PLC dominates the complex interactions at low temperatures in $Sr_3Ir_2O_7$. Note that, for the temperature dependence, a deviation between the AFM order parameter $M_{AF}^2$ and the amplitude $A_\Omega$, frequency softening $\Delta\nu$, lattice contraction $\Delta c$ exists, especially in the range of 100-280 K as shown in Figs. 3(a)&(c) and Fig. 4(c). We attribute it to possible other interactions, such as orbital-lattice interaction or charge-lattice interaction (see Fig. 3(d)), which we leave for future investigations. Anyhow, quantitatively, here the major interaction at low temperatures is still the PLC.

Interestingly, the strong PLC we find in $Sr_3Ir_2O_7$ frequently echoes its counterpart (strong SLC) in high temperature superconducting cuprates [35,36], or undoped cuprates [37], or even single crystal CuO [28]. Our findings here reveal possibly another novel shared similarity—strong PLC/SLC—between iridates and cuprates. So far, several analogies between cuprates and iridates have been found, for example, they both are $J(S)=1/2$ systems, Mott insulators at low temperature, full of rich electronic states, AFM in the parent material, undergo phase transition induced by doping, *etc*. However, these are mostly properties related to a single degree of freedom. Those related to *interactions* between different degrees of freedom are rarely reported. As interactions among charge, lattice, spin, and orbit are pivotal to understanding and realizing high temperature superconductivity, our identifying the *interaction*, strong PLC, constitutes a new bridge between iridates and cuprates. This shared similarity may shed new light on further investigations of iridates. For example, based on the strong PLC, applying strains along the *c*-axis can directly engineer the magnetic properties, such as modulating $T_N$ or even suppressing the AFM ordering. Thus our investigation suggests a new way of inducing or tuning the various properties in strongly correlated iridates, including possible superconductivity.

## V. CONCLUSION

In summary, we report an unambiguous observation of the extremely strong PLC in a single crystal of $Sr_3Ir_2O_7$ through an ultrafast laser spectroscopy investigation of the coherent phonon. Below $T_N$, a PLC-induced phonon anomaly (in amplitude, frequency and lifetime) is directly observed, and negative thermal expansion is discovered for the first time, both of which are direct effects of the strong PLC. We obtain an exceptionally high PLC coefficient for $Sr_3Ir_2O_7$, $\lambda = 150 \pm 20$ cm$^{-1}$, which is the largest PLC or SLC coefficient value ever reported. We identify that PLC is a crucial interaction at low temperatures in $Sr_3Ir_2O_7$. These results clearly reveal the central role of PLC in the iridates and mark another novel property shared between the iridates and cuprates. Our investigation provides a clear fundamental physical picture of the PLC in $Sr_3Ir_2O_7$, and suggests a new method of tuning the magnetic, lattice and electronic properties in this strongly correlated system, including possible superconductivity.

## ACKNOWLEDGEMENTS

This work was supported by the National Key Research and Development Program of China (2017YFA0303603, 2016YFA0300303, 2017YFA0302901), the National Natural Science Foundation of China (11574383, 11774408, 11774399), the Strategic Priority Research Program of CAS (XDB30000000), the External Cooperation Program of Chinese Academy of Sciences (GJHZ1826) and the CAS Interdisciplinary Innovation Team.

## APPENDIX A: SCHEMATIC LATTICE STRUCTURE AND ELECTRONIC ENERGY BAND DIAGRAM

The crystal structure of the bilayer perovskite iridate $Sr_3Ir_2O_7$ is shown in Fig. 5(a). It has a tetragonal unit cell ($a$ = 3.9 Å and $c$ = 20.9 Å) in space group $I4/mmm$. The $IrO_6$ octahedral bilayers are connected by the shared apical $O^{2-}$ ions. The SOC-induced pseudospins ($J$ = 1/2) on the $Ir^{4+}$ ions are aligned antiparallel along the $c$-axis. A schematic diagram of the electronic energy levels and bands is shown in Fig. 5(b). The crystal field splits the $5d$ orbitals into $t_{2g}$ orbitals and $e_g$ orbitals [38]. The strong SOC, which has a magnitude close to that of the crystal field [38], further splits the t$_{2g}$ Ir-5$d$ band into a lower $J_{eff}$ = 3/2 band and a higher $J_{eff}$ = 1/2 band [3]. Moreover, electron correlations further split the $J_{eff}$ = 1/2 band into a lower and an upper Hubbard band, respectively, with a bandgap of 130 meV [39]. The Fermi level is also marked in Fig. 5(b).

## APPENDIX B: SAMPLE GROWTH

Our single crystalline Sr$_3$Ir$_2$O$_7$ is grown by the flux method using SrCl$_2$•6H$_2$O as the solvent. The IrO$_2$ (99.99%), SrCO$_3$ (99.99%), and SrCl$_2$•6H$_2$O (99.9%) powders are thoroughly mixed with a molar ratio of 2:3:10, and then put into a platinum crucible, covered by a lid. The materials are heated to 1100 °C and kept for 24 hours. Then the melt is cooled to 800 °C at a rate of 4 °C/hr, followed by quenching in the furnace. Finally, the crystals are obtained by removing the flux with distilled water. The mass ratio between the product Sr$_3$Ir$_2$O$_7$ and reactant SrCl$_2$•6H$_2$O is 1:10.

## APPENDIX C: XRD AND RAMAN CHARACTERIZATION OF Sr$_3$Ir$_2$O$_7$ SINGLE CRYSTAL

The XRD pattern of a Sr$_3$Ir$_2$O$_7$ single crystal at 300 K is shown in Fig. 6(a). The regularly-spaced peaks correspond to the (0 0 n) reflections of Sr$_3$Ir$_2$O$_7$. The XRD characterization results shown here also indicate that, in the ultrafast experiment, the incident laser beam is perpendicular to the (0 0 1) planes.

A Raman scattering spectrum with 532 nm laser excitation at 300 K is shown in Fig. 6(b). The low-energy $A_{1g}$ phonon mode with a frequency of 4.4 THz, shaded in red, is extensively investigated in the main text.

## APPENDIX D: OBTAINING THE PLC COEFFICIENT $\lambda$ AND $\frac{\partial^2 J_c}{\partial u^2}$

As introduced in the main text, pseudospins in Sr$_3$Ir$_2$O$_7$ are anti-aligned along the $c$-axis. In Heisenberg's nearest-neighbor interaction picture, the Hamiltonian corresponding to the pseudospin interaction is

$$H_s = J_c \sum_i <J_i \cdot J_{i+1}>, \qquad (4)$$

where $J_c$ is the superexchange coupling energy between two nearest Ir$^{4+}$ ions and $J_i$ is the pseudospin of the $i$th Ir$^{4+}$ ion. Considering a pair of Ir$^{4+}$ ions, in the quantum harmonic oscillator picture, the Hamiltonian for the $A_{1g}$ phonon without PLC can be written as

$$H_{phonon} = \frac{P^2}{2\mu} + \frac{1}{2}k(\Delta u)^2, \qquad (5)$$

where $P$ is the momentum operator for the Ir$^{4+}$ ion, $\mu$ is the reduced mass (one half of the mass of an iridium atom), $k$ is the spring force constant between the two nearest anti-phase vibrating Ir$^{4+}$ ions and $\Delta u$ is the deviation from the equilibrium distance between the two Ir$^{4+}$ ions. When additionally considering the PLC, the equivalent spring force constant can be written as $k'$ and the Hamiltonian can be written as

$$H = H_{phonon} + H_{pseudospin} = \frac{P^2}{2\mu} + \frac{1}{2}k(\Delta u)^2 + NJ_c <J_i \cdot J_{i+1}> = \frac{P^2}{2\mu} + \frac{1}{2}k'(\Delta u)^2, \qquad (6)$$

where $N$ is the number of magnetic ion pairs in each $A_{1g}$ phonon (here $N = 1$). The $k'$ can be obtained by taking the second derivative of the potential term in Eq. (6):

$$k' = k + N \frac{\partial^2 J_c}{\partial u^2} <J_i \cdot J_{i+1}>, \qquad (7)$$

The oscillation frequencies with and without PLC can be obtained by $\Omega = \sqrt{k/\mu}$ and $\Omega' = \sqrt{k'/\mu}$. Thus, Eq. (7) can be written as

$$\Omega'^2 = \Omega^2 + (N/\mu)\frac{\partial^2 J_c}{\partial u^2} <J_i \cdot J_{i+1}>. \qquad (8)$$

Omitting the high order terms, the pseudospin-induced frequency anomaly

$\Delta\Omega = \Omega' - \Omega$ can be explicitly written as

$$\Delta\Omega = \frac{1}{2\Omega}(N/\mu)\frac{\partial^2 J_c}{\partial u^2}<J_i \cdot J_{i+1}>. \tag{9}$$

In the molecular field approximation [40], we have

$$<J_i \cdot J_{i+1}> \approx -|<J_i \cdot J_i>| = -\left(\frac{M_{AF}}{2\mu_B}\right)^2. \tag{10}$$

Substituting Eq. (10) into Eq. (9), we obtain

$$\Delta\Omega = -\frac{N}{8\mu\Omega}\frac{\partial^2 J_c}{\partial u^2}\left(\frac{M_{AF}}{\mu_B}\right)^2. \tag{11}$$

The PLC strength can be characterized by a constant $\lambda$ defined as follows [28,31]

$$\frac{\Delta\Omega}{2\pi} = \lambda <J_i \cdot J_{i+1}>, \tag{12}$$

where $\lambda$ is the conventionally defined spin-phonon coupling coefficient. As such,

$$\lambda = \frac{\Delta\Omega}{2\pi <J_i \cdot J_{i+1}>} = \Delta\nu/<J_i \cdot J_{i+1}> = -4\Delta\nu\mu_B^2/M_{AF}^2. \tag{13}$$

We interpret the maximum value of $M_{AF}^2$ from Ref. [24], which is $0.133\mu_B^2$. From Fig. 3(c), the maximum frequency deviation measured is $\Delta\nu = -5.0 \pm 0.6$ cm$^{-1}$. From Eq. (13), it is straightforward to obtain $\lambda = 150 \pm 20$ cm$^{-1}$ for Sr$_3$Ir$_2$O$_7$, which is larger than that of CuO [28]. Table 1 shows that Sr$_3$Ir$_2$O$_7$ has the largest PLC coefficient $\lambda$ among all the reported materials [27,31].

Furthermore, from Fig. 3(b), we obtain $\Omega/2\pi = 147$ cm$^{-1}$ at 0 K, from which we calculate using Eq. (11) that $\partial^2 J_c/\partial u^2 = 252 \pm 34$ N/m (*i.e.*, $1160 \pm 154$ mRy/Å$^2$). This value is two orders of magnitude larger than those of 3*d* transition metal oxides. For example, it is reported that $\partial^2 J_{xz}/\partial u_{stret}^2 = 16$ mRy/Å$^2$ for LaMnO$_3$ [27].

**APPENDIX E: RELATIVE INTENSITIES OF THE TWO $A_{1g}$ MODES UNDER**

**DIFFERENT EXCITATION WAVELENGTHS**

Notice that in our ultrafast generation and detection of the coherent phonon, the intensity of the low-energy $A_{1g}$ mode (142 cm$^{-1}$) is much higher than that of the high-energy $A_{1g}$ mode (183 cm$^{-1}$), as shown in the two insets of Fig. 1. However, comparable intensities of the two $A_{1g}$ peaks were observed in a Raman scattering experiment, using an excitation laser beam with the wavelength of 633 nm [7]. We unambiguously attribute this seemingly discrepancy to the different excitation laser wavelength. To see this, we summarize the reported results of phonon investigations in Fig. 7, where Fig. 7(a) is retrieved from Ref. [41], Fig. 7(b) is retrieved from Ref. [7], and Fig. 7(c) is our data shown in Fig. 1 of the main text. It can be seen clearly that, when the excitation laser wavelength increases, the relative intensities of the two $A_{1g}$ peaks change progressively and prominently: the intensity of the low energy $A_{1g}$ peak enhances and that of the high energy $A_{1g}$ peak decreases. Such dramatic change explains why we see different major phonon modes in these investigations. We contemplate that it is related to the details of the electronic band structure of Sr$_3$Ir$_2$O$_7$, which is worthy of further investigation.

*Structural Distortion-Induced Magnetoelastic Locking in $Sr_2IrO_4$ Revealed through Nonlinear Optical Harmonic Generation,* Phys. Rev. Lett. **114**, 096404 (2015).

[10] Y. C. Tian, W. H. Zhang, F. S. Li, Y. L. Wu, Q. Wu, F. Sun, G. Y. Zhou, L. Wang, X. Ma, Q.-K. Xue, and Jimin Zhao, *Ultrafast Dynamics Evidence of High Temperature Superconductivity in Single Unit Cell FeSe on $SrTiO_3$,* Phys. Rev. Lett. **116**, 107001 (2016).

[11] C. Aku-Leh, Jimin Zhao, R. Merlin, J. Menéndez, and M. Cardona, *Long-Lived Optical Phonons in ZnO Studied with Impulsive Stimulated Raman Scattering*, Phys. Rev. B **71**, 205211 (2005).

[12] F. Sun, Q. Wu, Y. L. Wu, H. Zhao, C. J. Yi, Y. C. Tian, H. W. Liu, Y. G. Shi, H. Ding, X. Dai, and P. Richard, *Coherent Helix Vacancy Phonon and Its Ultrafast Dynamics Waning in Topological Dirac Semimetal $Cd_3As_2$,* Jimin Zhao, Phys. Rev. B **95**, 235108 (2017).

[13] Jimin Zhao, A. V. Bragas, D. J. Lockwood, and R. Merlin, *Magnon Squeezing in an Antiferromagnet: Reducing the Spin Noise*, Phys. Rev. Lett. **93**, 107203 (2004).

[14] Jimin Zhao, A. V. Bragas, R. Merlin, and D. J. Lockwood, *Magnon Squeezing in Antiferromagnetic $MnF_2$ and $FeF_2$,* Phys. Rev. B **73**, 184434 (2006).

[15] D. Hsieh, F. Mahmood, D. H. Torchinsky, G. Cao, and N. Gedik, *Observation of a Metal-to-Insulator Transition with Both Mott-Hubbard and Slater Characteristics in $Sr_2IrO_4$ from Time-Resolved Photocarrier Dynamics*, Phys. Rev. B **86**, 035128 (2012).

[16] J. P. Hinton, J. D. Koralek, G. Yu, E. M. Motoyama, Y. M. Lu, A. Vishwanath, M. Greven, and J. Orenstein, *Time-Resolved Optical Reflectivity of the Electron-Doped $Nd_{2-x}Ce_xCuO_{4+\delta}$ Cuprate Superconductor: Evidence for Interplay Between Competing Orders*, Phys. Rev. Lett. **110**, 217002 (2013).

[17] Y. L. Wu, Q. Wu, F. Sun, C. Cheng, S. Meng, and Jimin Zhao, *Emergence of Electron Coherence and Two-Color All-Optical Switching in $MoS_2$ Based on Spatial Self-Phase Modulation*, Proc. Natl. Acad. Sci. U. S. A **112**, 11800 (2015).

[18] M. Mitrano, A. Cantaluppi, D. Nicoletti, S. Kaiser, A. Perucchi, S. Lupi, P. Di Pietro, D. Pontiroli, M. Riccò, S. R. Clark, D. Jaksch, and A. Cavalleri, *Possible Light-Induced Superconductivity in $K_3C_{60}$ at High Temperature*, Nature **530**, 461 (2016).

[19] J. Orenstein, *Ultrafast Spectroscopy of Quantum Materials*, Physics Today 65, 44 (2012).

[20] H. J. Park, C. H. Sohn，D. W. Jeong, G. Cao, K. W. Kim, S. J. Moon, H. Jin, D.-Y. Cho, and

**Figures and Captions**

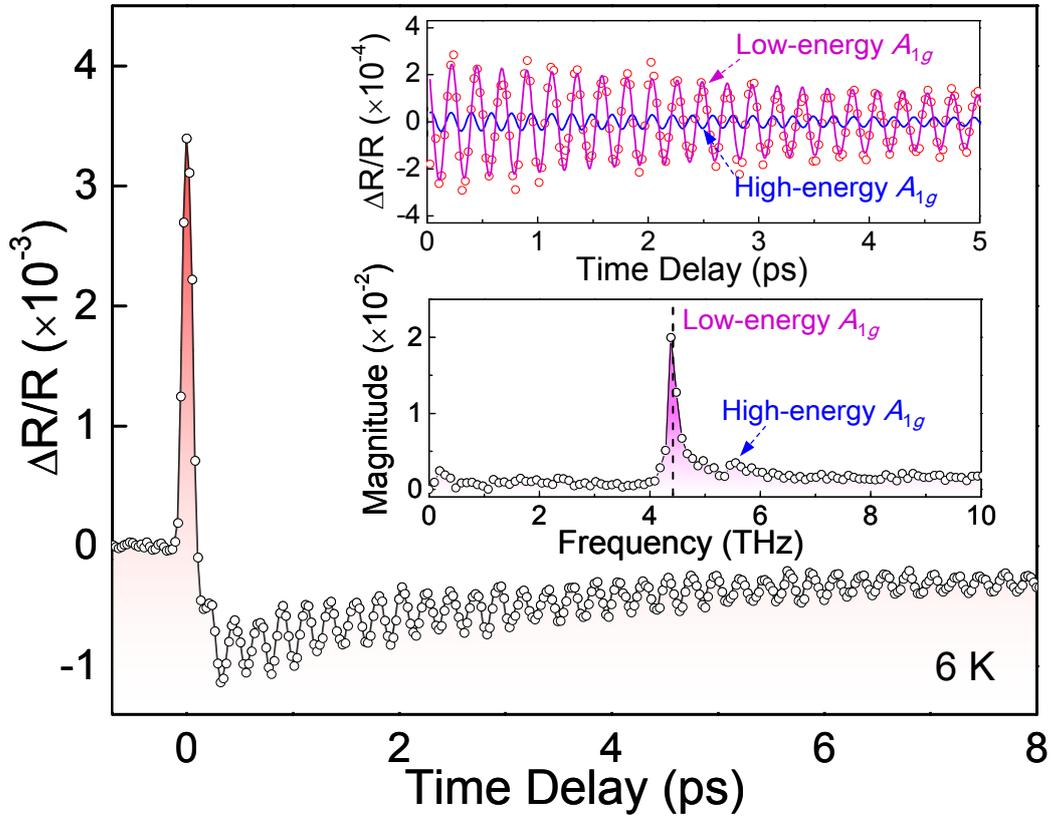

FIG. 1. Time-resolved ultrafast dynamics of $Sr_3Ir_2O_7$ at 6 K. The periodic oscillations superimposed on the electronic dynamics are the coherent optical phonons. Upper inset: coherent phonon oscillation after subtracting the electronic background, with the red (blue) curve being the damped waveform fit of the lower- (higher-) frequency $A_{1g}$ phonon mode. Lower inset: fast Fourier transformation of the experimental result in the upper inset.

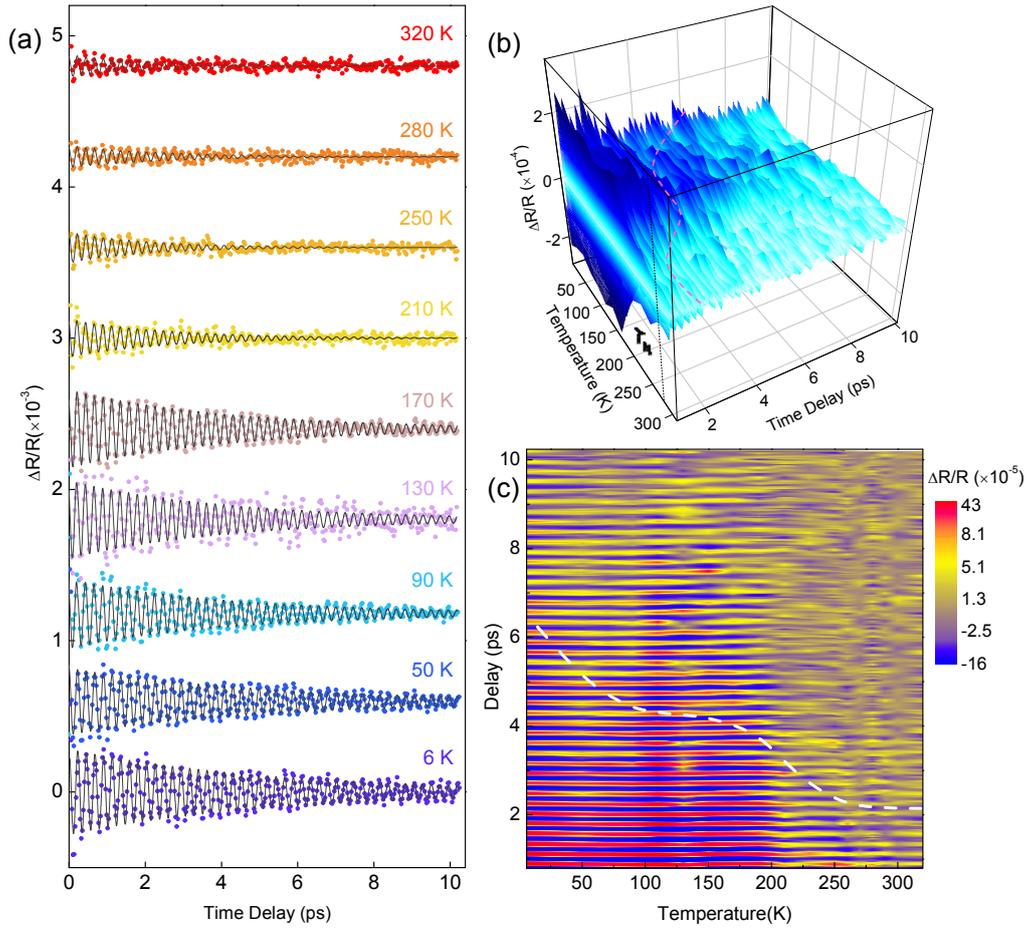

FIG. 2. (a) Time-resolved coherent phonon at several typical temperatures. The traces are offset for clarity. The three-dimensional (b) and two-dimensional (c) colormaps of the temperature dependence of the coherent phonons. The dashed curves represent the lifetime of the coherent phonon.

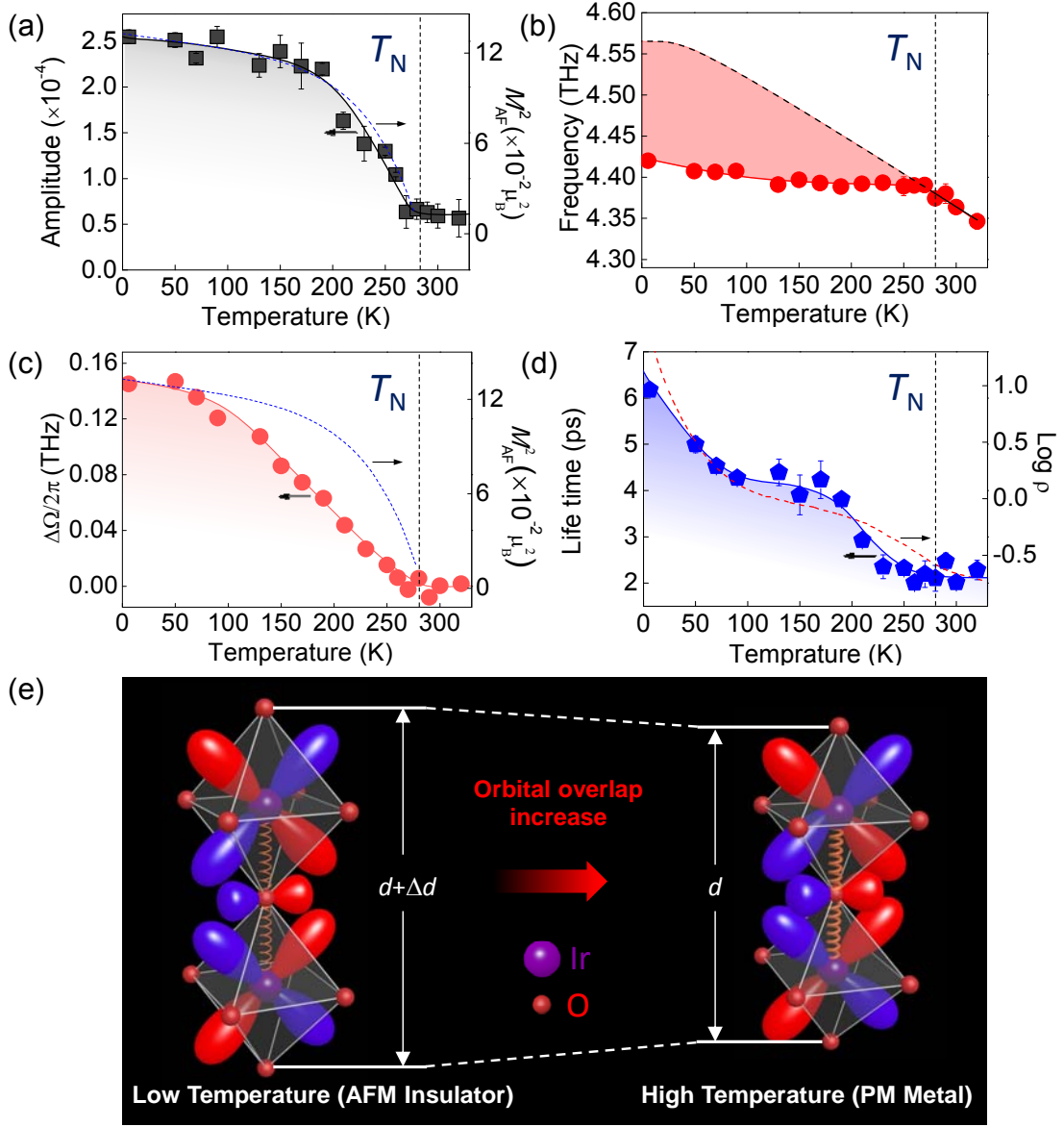

FIG. 3. Temperature dependence of the coherent $A_{1g}$ phonon. (a) Amplitude (black squares). Blue dashed curve: $M_{AF}^2$ adapted from Ref. [24]. (b) Frequency (red dots). Black dashed curve: expected frequency by assuming conventional anharmonic phonon decay. (c) Frequency deviation $\Delta v$ (red dots) assuming conventional anharmonic phonon decay. This value corresponds to the shaded area in (b). Blue dashed curve: $M_{AF}^2$ adapted from Ref. [24]. (d) Lifetime (blue pentagons). Red dashed curve: log ρ adapted from Ref. [32], with ρ being the resistivity. (e) Schematic illustration of the AFM pseudospin-induced enhancement of the interlayer spacing (i.e., PLC-induced anisotropic negative thermal expansion) and reduction of the orbital overlap. Solid curves in (a)-(d): visual guides. The dashed vertical lines mark $T_N$.

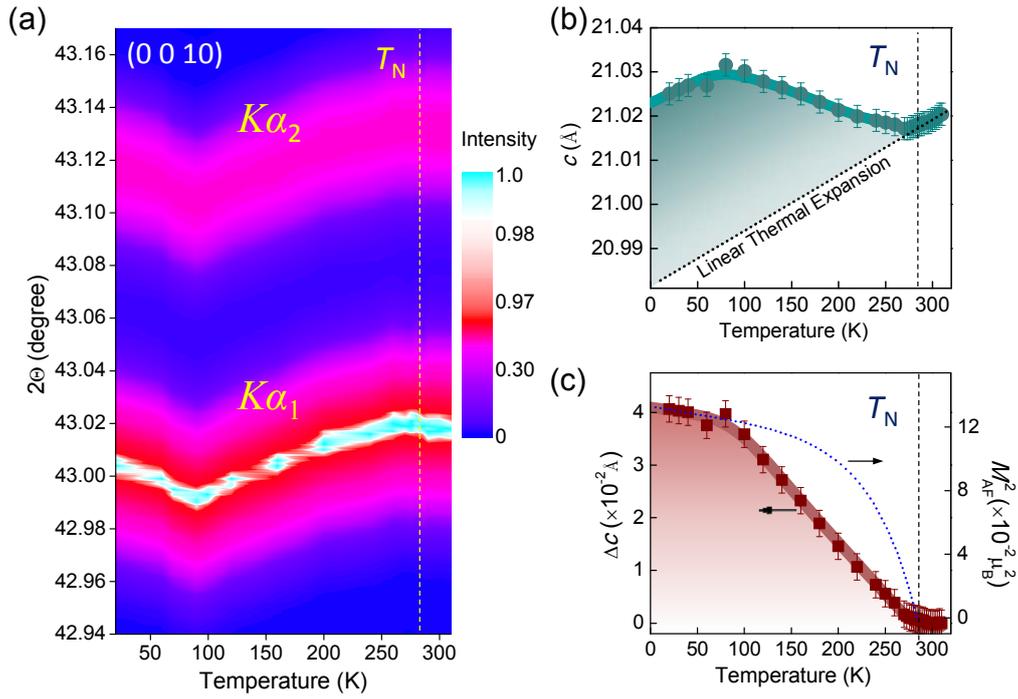

FIG. 4. Temperature-dependent single crystal XRD measurement. (a) Peak positions and intensities of the (0 0 10) Kα$_1$ and Kα$_2$ diffraction lines. (b) Lattice constant $c$ derived from the XRD results in (a) (dark green dots). Black dotted line: extrapolated (from results above $T_N$) linear thermal expansion. (c) Deviation (brown squares) $\Delta c$ in lattice constant from the conventional expectation of linear thermal expansion. This value corresponds to the dark green shaded area in (b). Blue dashed curve: $M_{AF}^2$ adapted from Ref. [24]. Dashed vertical lines mark $T_N$.

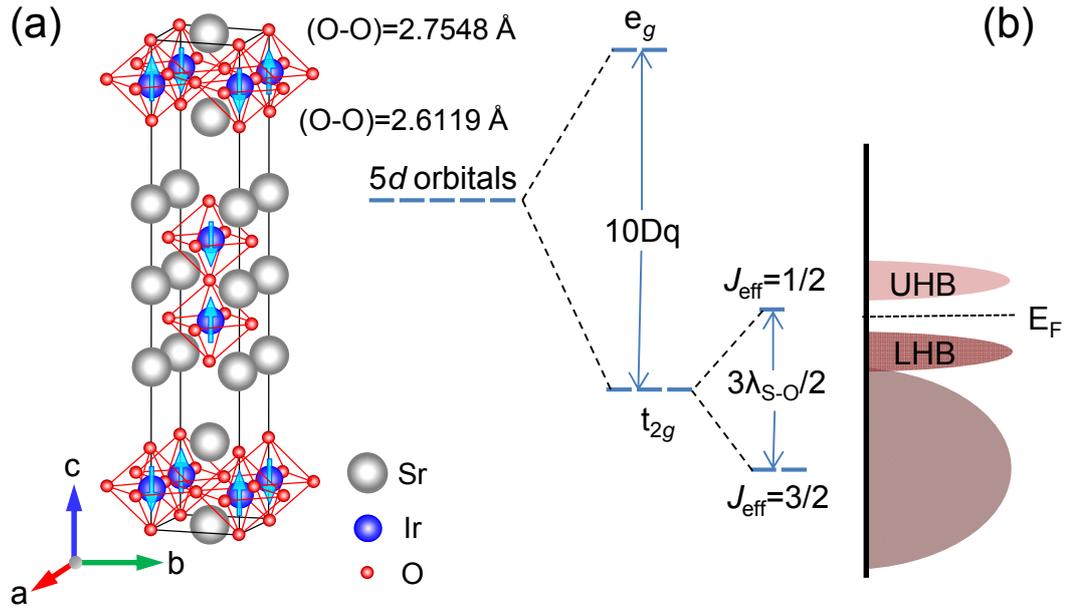

FiG. 5. Crystal and electronic structures. (a) Schematic crystal structure. The two neighboring IrO$_6$ layers are connected by the bridging O$^{2-}$ ions. The blue arrows represent the directions of the pseudospins on the Ir$^{4+}$ ions. (b) Schematic energy level and band diagram of the Ir$^{4+}$ ions, where crystal field splitting (10Dq), SOC (3$\lambda_{SO}$/2) and electron correlation all result in energy separations. UHB: upper Hubbard band. LHB: lower Hubbard band.

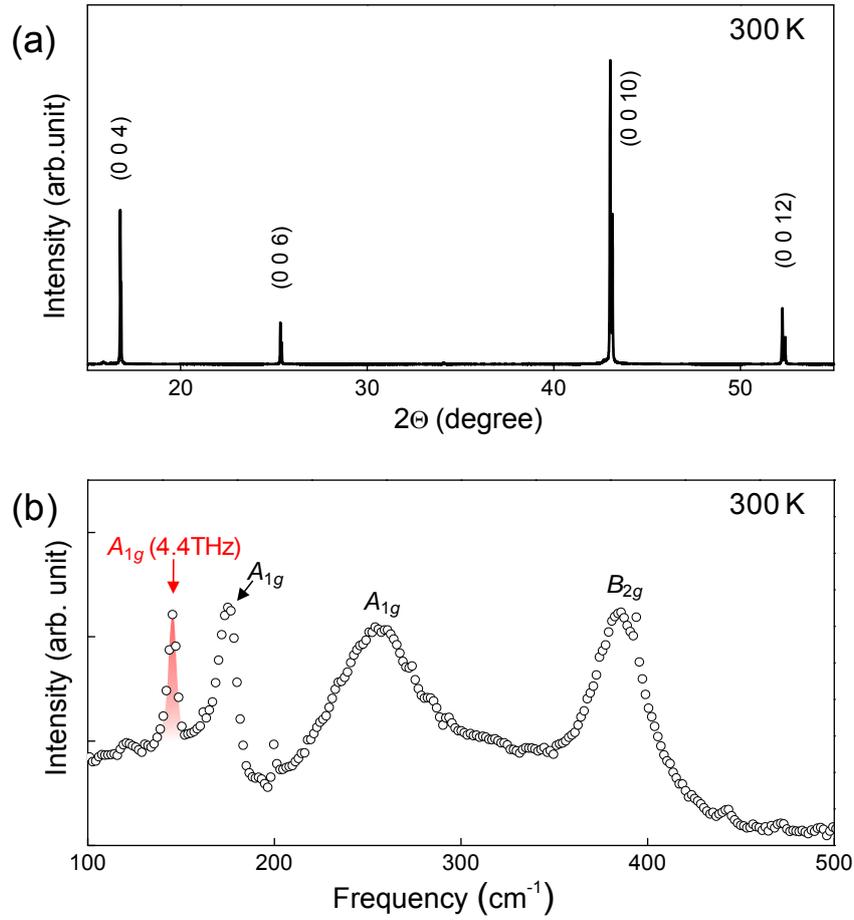

FIG. 6. Characterization of the Sr$_3$Ir$_2$O$_7$ single crystal at 300 K. (a) XRD pattern under Cu $K\alpha_{1,2}$ radiation, with the peaks denoted by index (0 0 n). These peaks are all double peaks, of which the weaker ones are caused by the $K\alpha_2$ radiation. (b) Raman spectrum under 532 nm laser excitation. The red shaded peak represents the $A_{1g}$ phonon mode corresponding to the out-of-plane IrO$_2$ layer breathing.

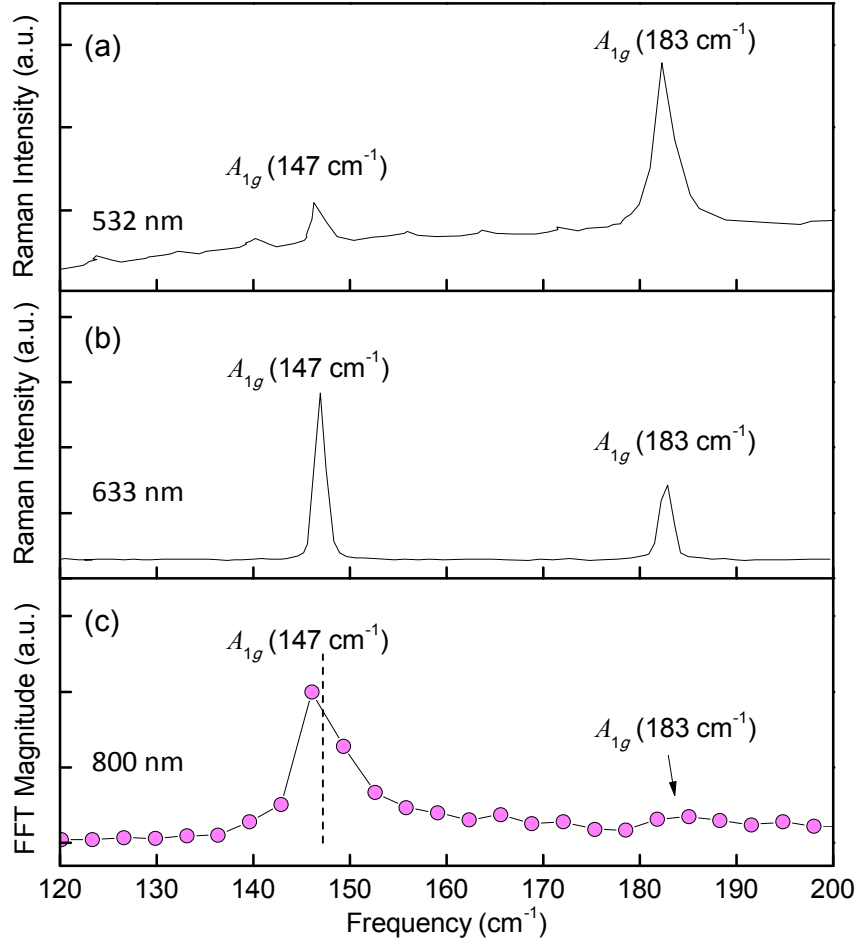

FIG. 7. The intensities of the two $A_{1g}$ modes at three different excitation laser wavelengths at low temperatures. (a) Raman scattering under laser excitation with a wavelength of 532 nm (5 K), whereby the data is adapted and retrieved from Ref. [41]. (b) Raman scattering under laser excitation with a wavelength of 633 nm (40 K), whereby the data is adapted and retrieved from Ref. [7]. (c) Coherent phonon generated in our ultrafast pump-probe experiment with a laser wavelength of 800 nm (6 K); the dashed line marks the frequency (147.3 cm$^{-1}$, or 4.420 THz) fitted by Eq. (1) in the main text.

TABLE 1. The PLC (or SLC) coefficient $\lambda$ for different compounds.

|  | $\Delta\nu$ (cm$^{-1}$) | $M_{sublattice}$ ($\mu_B$) | $|\lambda|$ (cm$^{-1}$) |
| --- | --- | --- | --- |
| Sr$_3$Ir$_2$O$_7$ | -5.0 ± 0.6 | 0.36 | 150 ± 20 |
| CuO[28,30] | 10 | 0.69 | 50 |
| LaMnO$_3$[27] | -8 | 3.65 | 2.4 |
| PrMnO$_3$[29] | -6.7 | 3.50 | 2.2 |
| NdMnO$_3$[29] | -5.2 | 3.22 | 2.0 |
| MnF$_2$[31] | -1.3~2.5 | 5 | 0.2~0.4 |